\documentclass[twocolumn]{svjour3}
\usepackage{graphics}
\usepackage{amssymb}
\usepackage{braket}
\journalname{Granular Matter}
\usepackage{epstopdf} 
\usepackage[usenames]{color}
\usepackage{hyperref}
\usepackage[utf8]{inputenc}
\usepackage[T1]{fontenc}
\usepackage[version=4]{mhchem}
\usepackage{soul}

\begin{document}

\title{Transparent Experiments: Releasing Data from Mechanical Tests on Three Dimensional Hydrogel Sphere Packings}

\author{Jonathan Bar\'es \and Nicolas Brodu \and Hu Zheng  \and Joshua A. Dijksman}

\institute{J. Bar\'es \at Laboratoire de M\'ecanique et G\'enie Civil, Universit\'e de Montpellier, CNRS, Montpellier, France 
           \and
           N. Brodu \at GeoStat, INRIA Bordeaux Sud-Ouest, Bordeaux, France
           \and
           H. Zheng \at Department of Physics and Center for Non-linear and Complex Systems, Duke University, Durham, North Carolina 27708, USA \and 
           J. A. Dijksman\\
           \email{joshua.dijksman@wur.nl}\at Physical Chemistry and Soft Matter, Wageningen University and Research, Stippeneng 4, 6708 WE Wageningen, Netherland}

\date{Received: date / Accepted: date}

\maketitle

%-/^\-%-/^\-%-/^\-%-/^\-%-/^\-%-/^\-%-/^\-%-/^\-%-/^\-%-/^\-%
\begin{abstract}
We describe here experiments on the mechanics of hydrogel particle packings from the Behringer lab, performed between 2012 and 2015.
These experiments quantify the evolution of all contact forces inside soft particle packings exposed to compression, shear, and the intrusion of a large intruder. The experimental set-ups and processes are presented and the data are concomitantly published in a repository~\cite{dryad3Drepo}.\\


\end{abstract}

%-/^\-%-/^\-%-/^\-%-/^\-%-/^\-%-/^\-%-/^\-%-/^\-%-/^\-%-/^\-%

\section{Introduction}\label{sec:intro}

Granular materials are ubiquitous; the many grains that compose them can be as small as confectionery sugar or as large as asteroids. Granular materials are relevant for disciplines ranging from the cement industry to pharmaceutical technology to the transport and production of food~\cite{rombauer1975joy},metals~\cite{gourlay2007dilatant}, ceramic bathroom tiles~\cite{cassani2001apparatus}, and even rechargeable batteries~\cite{Park2006}.
The microscopic interactions among grains are the essential foundation of macroscopic processes such as flow and deformation of these bulk granular materials.

A major challenge in understanding the mechanics of granular materials is that their behavior often depends on the microscopic interactions of the particles. As a consequence, it is imperative to perform experiments that can extract contact-level information. So far, most experimental measurements on granular micromechanics have been done via photoelasticimetry \cite{howell1999_prl,majmudar2005_nat,all2019_wiki} on two-dimensional model systems. Realistic granular materials, such as grains in the industries listed above, are three dimensional.
The need for experiments that can extract three dimensional (3D) contact level information from a granular material under realistic loading conditions is thus paramount. One imaging method that can be used to obtain 3D contact level information is refractive index matched scanning (RIMS), a technique that has been developed extensively over the past four decades~\cite{reviewRIMS,reviewsoftRIMS} and now finds applications ranging from fluid dynamics to biology, where it goes by the name of ``light sheet microscopy''~\cite{lightsheetmicro}.

Problematically, 3D imaging experiments are typically technically challenging, expensive and/or time consuming to carry out. They also provide tremendous amounts of data. Merely summarizing these data in concise graphs and tables as is typically done in scientific publications does not fully unlock its potential. Making raw experimental data available  would accelerate the field by allowing more detailed investigations and comparisons of experiments and theory. Data release would also allow the experimental data to be re-used for the calibration of numerical methods that aim to capture the mechanical behavior of granular materials. 

Here we provide the raw data from several years worth of effort on index matching experiments on soft hydrogel particle packings exposed to various loading conditions. All experiments have been performed in the Behringer lab at Duke University. Many of the data sets have not before been used, described explicitly in scientific publications or even analyzed. We will present a general overview of methods used to perform the imaging experiments. Due to data size limitations, we cannot provide the raw images for all experiments, but selected scans and smaller data sets are available. We provide particle-level information extracted from the 3D images, including contact forces and particle shapes, along with complementary boundary force measurements. We also provide a final working version of the code used to extract these microscopic features.

The general methodology of doing RIMS experiments on transparent macroscopic granular media, including soft particle packings, has been described extensively in previous work~\cite{reviewRIMS,reviewsoftRIMS}. We will therefore only describe the experimental protocols used for each experiment from which we release data.

%-/^\-%-/^\-%-/^\-%-/^\-%-/^\-%-/^\-%-/^\-%-/^\-%-/^\-%-/^\-%
\section{Experimental protocols}\label{sec:set-up}

\begin{figure}
\centering \resizebox{0.95\hsize}{!}{\includegraphics{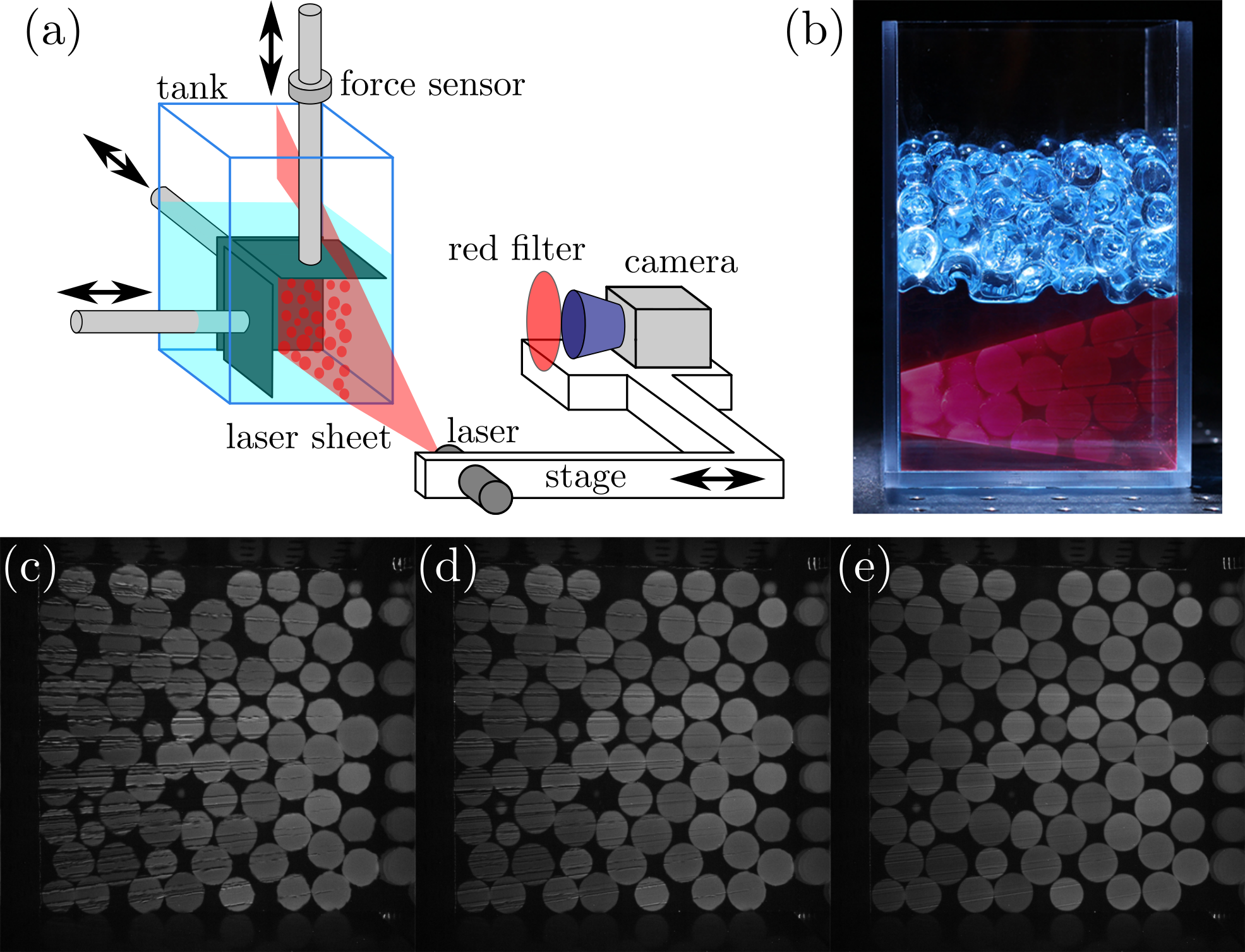}}
\caption{(a) Schematic view of the experimental set-up. A stage holding a camera equipped with a red filter and a laser creating a red light sheet is translated to scan a tank containing index-matched liquid and polyacrylamide beads that are loaded with three plates moving along three orthogonal axes.
(b) View of the polyacrylamide beads (top) in the index-matched liquid (bottom). The tank is partially illuminated with a laser sheet. Reprinted from Rev. Sci. Instr. \textbf{83}, 011301 (2012), with the permission of AIP Publishing. (c,d,e) Evolution (from left to right) of the view of a slice in the middle of the tank with increasing Polyvinylpyrrolidone (PVP) concentration, which is gradually added to index-match the liquid surrounding the particles.}
\label{fig1}
\end{figure} 

\begin{figure}
\centering \resizebox{0.85\hsize}{!}{\includegraphics{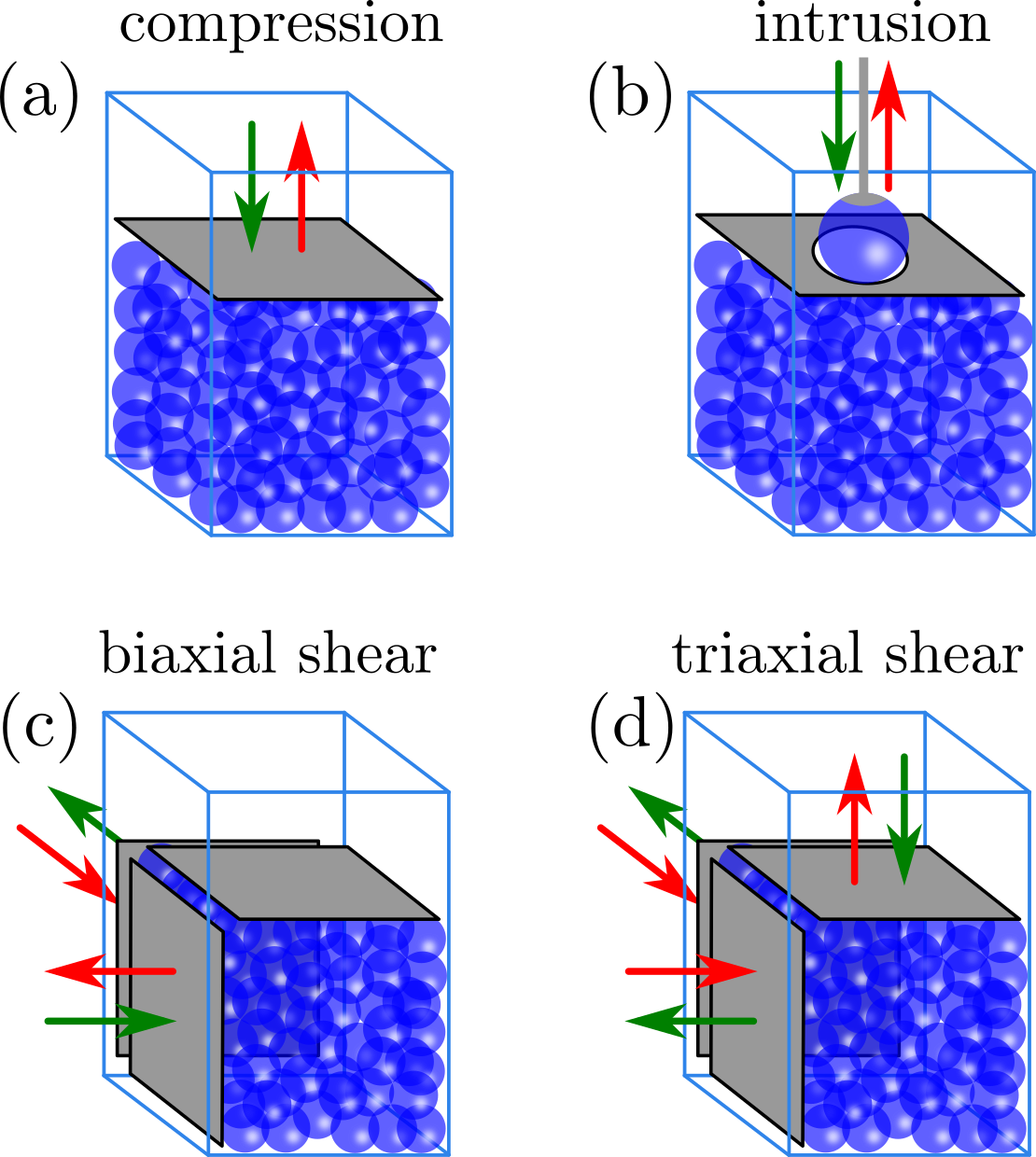}}
\caption{Schematic view of the different experimental loading mechanisms: (a) uniaxial compression of monodisperse and bidisperse grains; (b) intrusion of a large sphere in monodisperse grains; (c) isochoric biaxial shearing; (d) isochoric triaxial shearing. Green arrows stand for the displacement during the forward part of the loading while red arrows stand for the backward part.}
\label{fig2}
\end{figure}

A typical experimental setup is presented in fig.\ref{fig1}-a \cite{brodu2015_nc}. The system is composed of a transparent tank made of acrylic filled with hundreds of hydrogel particles (friction coefficient $\mu \sim 10^{-3}$) \cite{beads} that have been saturated with fluorescent dye. These particles are surrounded by a solution of water and polyvinylpyrrolidone (PVP) such that the index of refraction of the particles is well matched to the solution. As shown in fig.\ref{fig1}-b this reduces refraction of light at the hydrogel-water interface and improves optical access into the bulk of the packing. Also, since the density of the particles is very close to the one of the fluid, they experience an effective gravity of about $0.01$~$g$, where $g$ is the standard acceleration of gravity of $9.81~m/s^2$. Inside the tank the grains are mechanically loaded by three different moving plates. The displacement speed of one plate (called the leading plate) is exactly $1$~mm/s; for compression and intrusion (see Fig.~\ref{fig2}a-b) the other plates are fixed and for shear (Fig.~\ref{fig2}c-d) the speed of the other plates is slightly varied so that the displacement of all the plates begin and end together, keeping the system volume constant. The plates are made of acrylic and have a uniform pattern of holes drilled in them to minimize their viscous drag. The walls are oriented such that they keep the optical access free on both orthogonal sides of the packing. The system can thus be optically scanned with a camera whose image plane is parallel to a laser sheet. In all experiments series of $340$ to $360$ images are recorded by translating the stage holding the camera and laser, scanning the laser sheet through the sample to obtain stacks of $0.5$~mm thick image slices. 

As shown in fig.\ref{fig1}-b, we are able to image the entire inside of a three-dimensional (3D) system. The index matching of the particles makes them transparent in the surrounding solution unless the laser excites a cross-section of a particle, which fluoresces due to absorbed dye. A red longpass filter in front of the camera allows only the fluorescent light, with a longer wavelength that than of the laser light, to pass through to the camera,  preventing detection of occasional Reynolds scatter of laser light. In each experiment, the granular systems are quasistatically loaded; after each loading step, the system is scanned following a pause of a few seconds to allow the system to relax. From these scans, which each take several minutes, the evolution of the particle shapes are measured. The interparticle friction is negligible, so the contact force network can be extracted using an algorithm assuming purely normal contact forces already described in a previous publication \cite{brodu2015_nc} and available in the provided source code~\cite{dryad3Drepo}.

For each experimental protocol, during the whole experiment, a force sensor records the evolution of the vertical stress applied to the top of the sample (Loadstar RSB4-005M-A, $\pm$ 1 gram resolution; $\pm$ 0.02\% repeatability) via a 16-bit A/D converter. 

The swelling characteristics of the hydrogel particles is very sensitive to temperature. To avoid thermal dilation of the particles during the experiment, the whole set-up is contained in an insulated chamber where temperature is regulated and recorded at the top and bottom by thermocouples. Particles~\cite{beads} are grown $24$~h before each experiment at the operating temperature ($24^{\circ}$C) in deionized water (DI) and a ratio of Nile Blue dye corresponding with $20$~$\mu$L of solution per gram of dry hydrogel. Once fully grown, the particles are immersed in the set-up and surrounded by DI water. Strongly eccentric particles sometimes exist in a batch of grown particles; these are removed after manual optical assessment of sphericity. A saturated stock solution of PVP is added drop by drop to the container with particles, where it is mixed by continuously circulating the water-PVP mixture with a peristaltic pump connected with tubing to the container area outside the scanning volume to which the particles are confined. Due to porosity of the confining walls and the large amount of space between container walls and moving walls, the circulation method also effectively mixes the scanning volume. During the mixing, a slice in the middle of the sample is imaged. As presented in fig.\ref{fig1}-c to e, these slices get sharper when the index matching gets better. PVP addition and mixing is stopped when a clear image is obtained. The procedure to determine the moment at which index matching is good enough is hard to describe quantitatively; for more information, please see a review article~\cite{reviewRIMS}.

%-/^\-%-/^\-%-/^\-%
\subsection{Uniaxial Compression}

As shown in fig.\ref{fig2}, many different experimental protocols have been adapted to the set-up previously described. Here we describe the experiments about which previous works have appeared \cite{brodu2015_nc}. They are summarized in table~\ref{table:overview}. Fig.\ref{fig2}-a presents how a packing of $514$ slightly monodisperse hydrogel beads \cite{beads} of mean diameter $2.1$~cm are vertically compressed. The experimental protocol consists of a series of $20$ uniaxial compressions, each followed by a corresponding 20-step expansion to the original boundary configuration. Each cycle consists of a compression phase imposing a total strain up to $13.4$\% of the initial height, followed by a decompression phase. A full cycle is carried out in $60$ quasi-static steps of $1$~mm each. This experiment is referred to as ``Comp20''. Note that we verified that particle size in the ``Comp20'' experiment is constant to within 1.5\% of the total volume occupied of the particles as extracted with the shape extraction technique referenced below. We also completed an experiment on the same particles for compression $10$ cycles; this experiment is named ``Comp10''. 

To calibrate the relaxation dynamics of a packing under compression, we performed a compression experiment consisting of a single compressive phase, after which packing dynamics is recorded at constant finite strain. These two experiment are called ``CompRelax1/2''.  

A fourth uniaxial compression experiment has been carried out with a bidisperse packing of $1573$ small and $348$ medium sized beads. Medium sized beads are similar to the ones mentioned above, while the mean diameter of the smaller ones is $1.16$~cm. The maximum strain is $10.2$\%; this strain is achieved in $60$ compression steps of $0.5$~mm. This experiment is named ``CompBiDisp''.

%-/^\-%-/^\-%-/^\-%
\subsection{Intrusion in dense packing}\label{sec:intru}

Here we describe the experiment presented in fig.\ref{fig2}-b. A large hydrogel sphere of diameter $6.3$~cm \cite{beads} is pushed back and forth into a packing of $823$ monodisperse (mean diameter $2.1$~cm) beads confined to a fixed rectangular box whose top has a hole slightly larger than the diameter of the intruder bead. The intruder is attached to a force sensor with a suction mechanism that holds the large bead. The large bead is pushed up and down in the granular packing for $20$ cycles. The displacement amplitude of the intruder is $3$~cm made in $60$ quasi-static step of $1$~mm each. This experiment is named ``Intrusion''.

%-/^\-%-/^\-%-/^\-%
\subsection{Isochoric 2D shear}

As shown in fig.\ref{fig2}-c the isochoric biaxial shear experiment consists in compressing a packing of $757$ monodisperse beads (mean diameter $2.1$~cm) in one horizontal direction while expanding it in the orthogonal direction. Orthogonal motion is adjusted such that the overall volume is kept constant: The horizontal confinement wall (which confines particles vertically) does not move. The amplitude displacement of loading walls is $1$~cm, which induces a $12$\% shear amplitude. The particle packing is sheared back and forth for $20$ cycles of $20$ steps of $0.5$~mm. This experiment is named ``BiaxShear''. 

%-/^\-%-/^\-%-/^\-%
\subsection{Isochoric 3D shear}

Here we describe the isochoric triaxial shear experiment presented in fig.\ref{fig2}-d. A packing of $882$ monodisperse beads (mean diameter $2.1$~cm) is compressed vertically while it is expanded in the two orthogonal horizontal directions, keeping the overall volume constant. The total displacement of the top wall is $1.7$~cm, which induces a $11.5$\% shear strain. The system is sheared back and forth for $25$ cycles of $34$ steps of $0.5$~mm. This experiment is named ``TriaxShear''.

\begin{table} [!h]
\begin{tabular*}{0.48\textwidth}{@{\extracolsep{\fill}}lllll}
\hline
Experiment name  & Type   & \#3D frames \\
	\hline  
       Comp20 & uniaxial compr. & 1200 \\
       Comp10 & uniaxial compr. & 600  \\ 
	   CompRelax & uniaxial compr. + relax. & 30 \\
	   CompRelax2 & uniaxial compr. + relax. & 30 \\
	   CompBiDisp & uniax. compr. bidisp. & 180\\
	   Intrusion & Object intrusion & 1200 \\
	   BiaxShear & 2 moving bound. & 400\\ 
	   TriaxShear & 3 moving bound. & 1700 \\
    \hline
\end{tabular*}
\caption{An overview of all experimental data sets described here and provided as data sets in the associated repository.}
\label{table:overview}
\end{table}

%-/^\-%-/^\-%-/^\-%-/^\-%-/^\-%-/^\-%-/^\-%-/^\-%-/^\-%-/^\-%

\section{Experimental results}\label{sec:result}

Data corresponding with experiments described in the previous section are compiled in a Dryad repository~\cite{dryad3Drepo}. An overview of these experiments is shown in Table~\ref{table:overview}. The data have some notable aspects:
\begin{itemize}
    \item It is possible to extract the moment of inertia tensor from the shape data in all experiments. 
    \item CompRelax reveals relaxation of the compression force, but not of the particle contacts. The second CompRelax experiment is a duplicate but with twice the scanning resolution, without image analysis results.
    \item BiaxShear is accompanied by a set of force measurements on a wide variety of biaxial shear tests with different amplitudes. There is no complementary image data, but a MATLAB loader is provided.
\end{itemize}

\begin{figure}
\centering \resizebox{0.95\hsize}{!}{\includegraphics{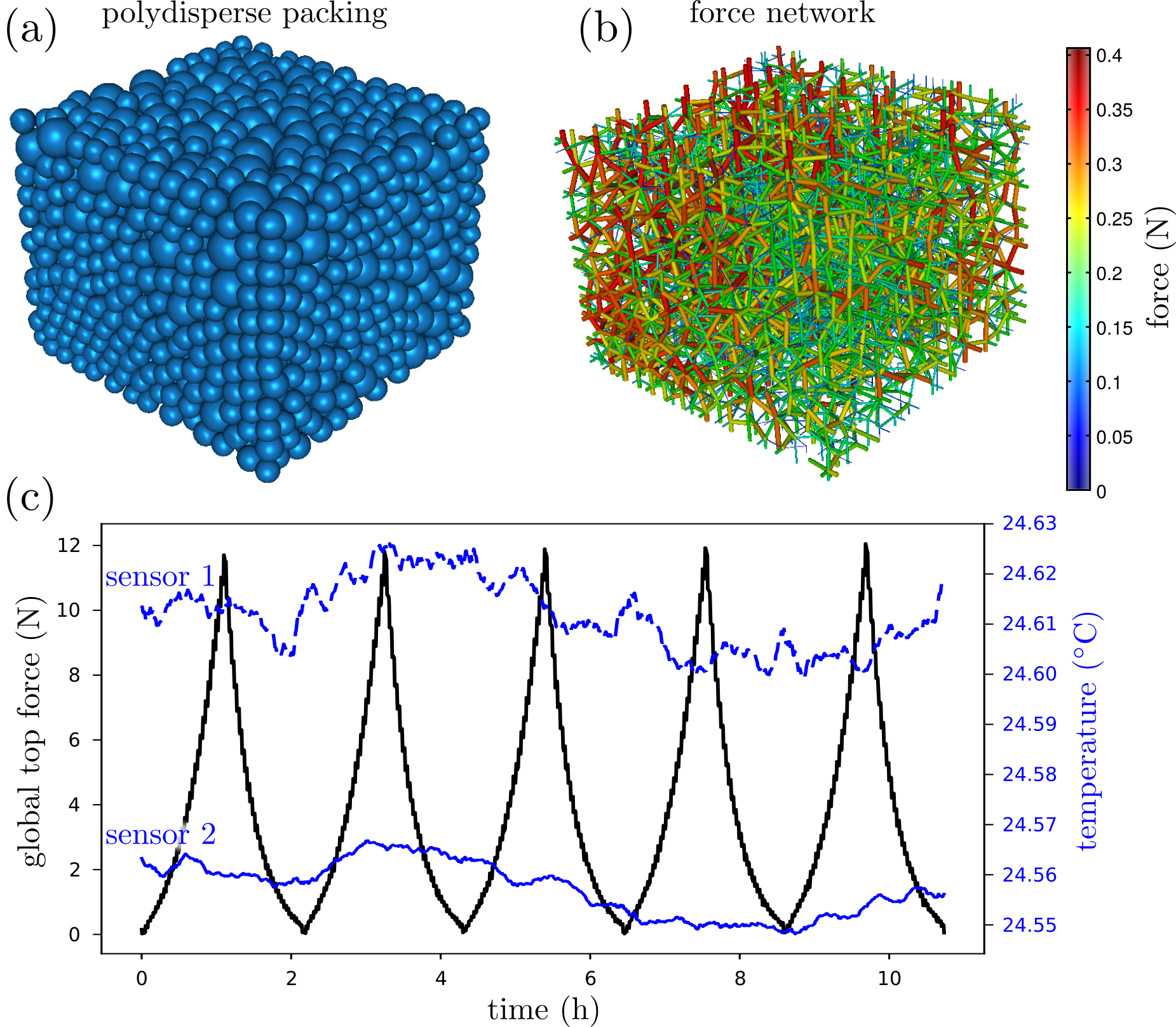}}
\caption{Results for the polydisperse compression experiment. (a) 3D reconstruction of the granular packing at $10$\% of strain, (b) corresponding force network, (c) evolution of the vertical force applied on the top wall during 5 cycles and of the temperature inside the insulated chamber (sensor 1 top, sensor 2 bottom).}
\label{fig3}
\end{figure} 

\section{Data Structure}\label{sec:datastruct} 

Each experiment, except for CompRelax2, has been post-processed with the image post-processing code outlined in \cite{reviewsoftRIMS}. The code is available in the repository as well. Generally speaking, for each experiment, the data provides the following information in the same format. Exceptions are described in the repository in human readable text files.  
\begin{itemize}
\item[+] Directory ``GlobalMeasure'' with: 
\item \verb|PlateMotion.txt|: x, y and z plate displacements for each step of a cycle.
\item \verb|Temperature.txt|: T1 and T2 as a function of time (see for example fig.\ref{fig3}-c).
\item \verb|TopForce.txt|: force as a function of time applied to the top plate or intruder (see for example fig.\ref{fig3}-c)
\item This directory may also include raw log files and MATLAB loader
\item[+] Directory ``LocalMeasure'' with: 
\item \verb|step_XXXX.txt.gz| The compressed text files in a YAML format~\cite{ben2005yaml} generated at the end of the image analysis code, with \verb|XXXX| the number of the loading step.
\item This directory may also include raw images and/or shape information files
\item[+] Directory ``Movie'' with:
\item \verb|Force.avi|: movie of the force network evolution during one cycle (see for example fig.\ref{fig3}-b).
\item \verb|Shape.avi|: movie of the grain evolution during one cycle (see for example fig.\ref{fig3}-a).
\item \verb|IndexMatching.avi|: evolution of a given slice when index matching
\item \verb|Scan.avi|: random scan movie.
\item \verb|InputParameter.txt|: a text file with experimental inputs.
\end{itemize}

The \verb|step_XXXX| data files contain the main data about each grain and each contact. They can easily be parsed and a MATLAB loader is provided. All units are SI except angles; they are in degrees. All position and vector values are given in 3D. Each file is divided in five main sections :
\begin{itemize}
\item \verb|stats|: Global statistics are provided for the whole packing. The most important are the total volume of the grains, the boundary forces, the number of grains and contacts. Derived statistics, such as the average number of contacts per grain, are also provided for convenience.
\item \verb|grains|: Each grain is given an ID and its characteristics are provided such as its center of mass, the displacement from the previous image, the grain volume and the list of contact IDs.
\item \verb|walls|: The 6 confinement boundaries are recovered from the images and a best fit plane is provided for each (in the form of a normal vector + offset).
\item \verb|contacts|: Each contact is also given an ID and its characteristics are provided. The main ones are the two grain or wall IDs that are in contact, the position of the contact, the force exerted by the first grain on the second, the computed 3D direction vector for this force and the area of the contact between the grains. 
\item \verb|shapes|: The surface of each grain is provided as a set of coefficients in a triangular spherical b-spline basis~\cite{he2004surface}. This is a basis of functions on the sphere giving, for each unit direction from the center of mass, the distance of the surface in that direction. The shape information provides complete surface structure data without having to resort to (truncated) spherical harmonics, with a resolution finer than the experimental voxel size due to the fit procedure. See Ref.~\cite{reviewsoftRIMS} for details. All grain information (including volume, contact area, etc) is computed from this analytic description. A visualization code is also provided. 
\end{itemize}

Not every package will have all elements included. Main exceptions are listed in Table~\ref{table:except}; otherwise the exceptions and data set details can be found in the data repository. 

\begin{table} [!h]
\begin{tabular*}{0.48\textwidth}{@{\extracolsep{\fill}}lllll}
\hline
Filename  & Description \\
	\hline  
	   \verb|PlateMotion.txt| & only for Comp20. \\
	   \verb|Temperature.txt| & provided through logfile.\\
	   \verb|TopForce.txt| &  provided through logfile.\\
	   \verb|step_XXXX.txt.gz| & not available in CompRelax2\\ 
    \hline
	   \verb|Force.avi| & N/A in Comp20/10/Relax/Relax2\\
	   \verb|Shape.avi| & N/A in Comp20/10/Relax/Relax2\\
	   \verb|IndexMatching.avi| & N/A in Comp20/10/Relax/Relax2\\
	   \verb|InputParameter.avi| & N/A in Comp20/10/Relax/Relax2\\
    \hline
\end{tabular*}
\caption{An overview of main exceptions to the data standard described.} 
\label{table:except}
\end{table}
%-/^\-%-/^\-%-/^\-%-/^\-%-/^\-%-/^\-%-/^\-%-/^\-%-/^\-%-/^\-%

\section{Discussion}\label{sec:disc}

Our aim with the release of these data sets is that these experimental data contribute to a better understanding of the multiscale physics of granular materials in general. How do we see this working out? One possibility is that this data release accelerates the integration of experimental calibration data in numerical models \cite{brodu2015_pre}, helping to overcome the limitations of existing calibration methods for granular materials \cite{paulick2015review,rackl2017calib}. Development of DEM simulations, which model individual particle dynamics in large systems, has led over the past few decades to successful modeling of some complex industrial processes~\cite{amarsid2017_pre} and even mixtures, such as suspensions with grains immersed in liquid. Due to this modeling success and the accessibility of the DEM approach, so-called ``digital twins'' are an important theme of many granular studies~\cite{rosen2015}. 
However, these increasingly complex computer methods now have many tuning parameters and thus require extensive experimental calibration to \emph{see} if they work at all, let alone \emph{improve} them for broader applicability. We believe that 3D imaging of granular materials has the potential to resolve these calibration issues to a large extent. It is therefore urgently necessary to develop advanced imaging tools that are combined with mechanical testing (such as rheology, uni/triaxial testing, adhesion testing) while characterizing DEM-relevant information such as grains contact forces, grain deformation and fluid interactions. Fortunately, 3D imaging methods such as X-ray and MRI have made huge leaps in the last decade \cite{richard2003_pre,huang2005_prl,sanfratello2007_gm,kudrolli2008}; development is still ongoing \cite{bares2017_epj}. It is now possible, for example, to 3D scan a suitcase in less than a second, an extremely useful development for airport security~\cite{Warnett_2016}. These developments should make it possible to arrive at the required advanced imaging tools. We hope that by releasing a set of data on model experiments that bridge the micro-macro gap~\cite{brodu2015_nc}, we assist in the convergence of the experimental and DEM communities to advance the understanding of granular materials in general.

\begin{acknowledgements}
We remember the late professor Bob Behringer for his kindness, for his ability to communicate with experts from a wide variety of science and engineering backgrounds, for his openness in scientific discussions and for providing a stimulating environment for every new generation of scientists passing through his laboratory. This work was partially funded by the EU Horizon 2020 MSCA ITN program CALIPER with grant number 812638. Funding was also provided by National Science Foundation (Grant No. 1206351), W. M. Keck Foundation, National
Aeronautics and Space Administration (Grant Nos. NNX10AU01G, NNX15AD38G) and DMR 1809762 and
ARO W911NF-18-1-0184. 
\end{acknowledgements}

\bibliographystyle{unsrt}
\bibliography{biblio}

\end{document}